\documentclass[letterpaper]{article} % DO NOT CHANGE THIS
\usepackage{aaai2026}
\usepackage{times}  % DO NOT CHANGE THIS
\usepackage{helvet}  % DO NOT CHANGE THIS
\usepackage{courier}  % DO NOT CHANGE THIS
\usepackage[hyphens]{url}  % DO NOT CHANGE THIS
\usepackage{graphicx} % DO NOT CHANGE THIS
\usepackage{natbib}  % DO NOT CHANGE THIS AND DO NOT ADD ANY OPTIONS TO IT
\usepackage{caption} % DO NOT CHANGE THIS AND DO NOT ADD ANY OPTIONS TO IT
\usepackage{algorithm}
\usepackage{algorithmic}
\usepackage{booktabs}
\usepackage{multirow}
\usepackage{amsmath}
\usepackage{tabularx}

\usepackage{newfloat}
\usepackage{listings}
\DeclareCaptionStyle{ruled}{labelfont=normalfont,labelsep=colon,strut=off} % DO NOT CHANGE THIS
\floatstyle{ruled}
\newfloat{listing}{tb}{lst}{}
\floatname{listing}{Listing}
\usepackage{enumitem}
\usepackage{multirow}
\usepackage{makecell}

\usepackage{todonotes}
\usepackage{subcaption}
\usepackage{xspace}

\author{
    Philipp Eibl\textsuperscript{\rm 1, 2}\equalcontrib,
    Erica Coppolillo\textsuperscript{\rm 1, 2, 3, 4}\equalcontrib, 
    Simone Mungari\textsuperscript{\rm 1, 2, 3, 4, 5}\equalcontrib and
    Luca Luceri\textsuperscript{\rm 1, 2}
}
\affiliations{
    \textsuperscript{\rm 1}University of Southern California, Los Angeles, California\\
    \textsuperscript{\rm 2} Information Sciences Institute (ISI), University of Southern California, Los Angeles, California\\
    \textsuperscript{\rm 3}University of Calabria, Rende, Italy\\
    \textsuperscript{\rm 4}ICAR-CNR, Rende, Italy\\
    \textsuperscript{\rm 5}Revelis s.r.l, Rende, Italy\\
}

\title{Is Grokipedia Right-Leaning? Comparing Political Framing in\\Wikipedia and Grokipedia on Controversial Topics}

\begin{document}

\maketitle

\begin{abstract}
Online encyclopedias are central to contemporary information infrastructures and have become focal points of debates over ideological bias. Wikipedia, in particular, has long been accused of left-leaning bias, while Grokipedia, an AI-generated encyclopedia launched by xAI, has been framed as a right-leaning alternative. This paper presents a comparative analysis of Wikipedia and Grokipedia on well-established politically contested topics. Specifically, we examine differences in semantic framing, political orientation, and content prioritization. We find that semantic similarity between the two platforms decays across article sections and diverges more strongly on controversial topics than on randomly sampled ones. Additionally, we show that both encyclopedias predominantly exhibit left-leaning framings, although Grokipedia exhibits a more bimodal distribution with increased prominence of right-leaning content. 
%Overall, our results challenge the portrayal of Grokipedia as an extreme right-leaning alternative.
% Online encyclopedias function as critical infrastructures within contemporary digital knowledge ecosystems. Wikipedia, in particular, is deeply embedded in search engines and digital assistants, shaping how information is accessed and interpreted at scale. At the same time, it has become the subject of recurring debates concerning ideological bias and the governance of knowledge. As an explicit alternative, xAI launched Grokipedia, an AI-generated online encyclopedia produced and maintained by a proprietary large language model.
% Motivated by accusations that Wikipedia promotes left-leaning framings, as well as by claims portraying Grokipedia as a more right-leaning alternative, this paper presents a comparative analysis of the two platforms focusing on well-established politically controversial topics. Using computational text analysis of article contents, we examine differences in semantic framing and assess whether systematic discrepancies in political orientation emerge across the two encyclopedias. Rather than adjudicating factual correctness, our analysis aims to evaluate whether, and to what extent, such systematic differences persist.
The experimental code is publicly available.\footnote{\url{https://anonymous.4open.science/r/grokipedia-FFC9}}
%Our findings contribute to ongoing discussions in social computing about algorithmic gatekeeping, bias, and the role of large language models in public knowledge infrastructures.
\end{abstract}

% \begin{links}
%   \link{Code}{https://anonymous.4open.science/r/grokipedia-FFC9}
% \end{links}
\section{Introduction}

Online encyclopedias are a core part of the contemporary information ecosystem.
Among them, Wikipedia has become a central piece of web infrastructure, widely embedded in search engines, digital assistants, and everyday information seeking.
At the same time, Wikipedia has long been the subject of debate about ideological bias, systemic under-representation, and the governance of knowledge more broadly~\cite{rozado2024wikipedia, greenstein2012wikipedia, hube2017bias, hube2018detecting, umarova2019partisanship}.
These concerns have motivated a series of alternative encyclopedic projects, including initiatives that are explicitly ideologically motivated~\cite{luyt2025conservapedia}.
%\todo[inline]{citare qui conservapedia senza menzionarla (magari citarla esplicitamente nei related)}, to commercial knowledge graphs integrated into search platforms~\cite{XXX}

In late 2025, xAI, the artificial intelligence company founded by Elon Musk, launched \emph{Grokipedia}, an AI-generated online encyclopedia positioned explicitly as a competitor to Wikipedia~\cite{theprint_grokipedia, pcmag_grokipedia, scmp_grokipedia}.
Unlike Wikipedia, where content is written and curated by a large community of volunteer editors, Grokipedia entries are generated and maintained by the Grok large language model~\cite{grok}, with users limited to suggesting corrections via a feedback interface.
Moreover, public statements around the launch frame Grokipedia as a way to ``purge propaganda'' and overcome alleged left-leaning bias in Wikipedia~\cite{wsj_wikipedia_bias, nytimes_grokipedia}.

Nevertheless, early studies on Grokipedia have raised concerns on several fronts, from replicating Wikipedia content with minimal changes~\cite{triedman2025did} to adopting non-trasparent sourcing~\cite{mehdizadeh2025epistemic, triedman2025did}. 
% These observations position Grokipedia not simply as another encyclopedia, but as a high-profile experiment in using a proprietary AI model to generate and gatekeep reference knowledge at scale.
% From the perspective of social computing and web science, Grokipedia raises at least two important questions.
% First, how does an AI-generated encyclopedia content \emph{semantically differ} compared to a community-edited one, on well-established controversial topics?
% %Ideological bias in Wikipedia has been extensively studied, but much less is known about how biases manifest when a single model, rather than a large and heterogeneous editor community, is responsible for producing content.
% Second, does the \textit{political orientation} of their framings on such topics significantly differ?
% Citations play a central role in Wikipedia’s governance and reliability, yet the extent to which AI-generated articles rely on different types of sources---including highly polarizing or low-credibility outlets---remains unclear.
% To the best of our knowledge, neither of these has been answered in the existing literature.
Motivated by these considerations, this paper presents a comparative analysis of Grokipedia and Wikipedia focused on well-established politically contested topics. This deliberate restriction enables a meaningful comparison under a political lens, as such topics are known to be highly polarized. Consequently, this setting allows us to statistically assess whether systematic political bias emerges across the two platforms. 
%We study both (1) political and ideological framings in article text and (2) the distribution and characteristics of cited sources, with particular attention to politically extreme or polarizing outlets.
%We underscore that our goal is not to adjudicate which platform is ``correct'', but to characterize how an AI-generated encyclopedia reconfigures the informational and political landscape relative to an established, community-governed one.
% Concretely, we structure our study around the following research questions:

% \begin{enumerate}[leftmargin=1cm]
%     \item[\textbf{RQ1}:] How do \textit{semantic framings} of controversial topics differ between Grokipedia and Wikipedia?
%     \item[\textbf{RQ2}:] Can we detect significant discrepancies in terms of \textit{political orientation}?
% \end{enumerate}

% \todo[inline]{ALTERNATIVE:}
Concretely, we make the following \textbf{contributions}:
\begin{itemize}[leftmargin=*]
    \item We provide a comparative analysis of the political orientation of Wikipedia and Grokipedia on a set of well-established controversial topics.
    \item Contrary to common narratives that portray Grokipedia as espousing extreme right-leaning views, we demonstrate that both platforms predominantly employ left-leaning framings overall.
    \item 
   Nevertheless, we identify decaying semantic similarity across articles and show that Grokipedia exhibits a stronger tendency to prioritize right-leaning content within article pages.   
\end{itemize}
To the best of our knowledge, this study represents the first systematic effort to examine political framing across Wikipedia and Grokipedia on controversial topics.

\section{Related Work}

With the recent emergence of Grokipedia, scholarly interest has intensified around systematic comparisons between this AI-driven encyclopedia and the long-established Wikipedia platform. 
Prior work has highlighted substantial differences in their epistemic foundations and content production practices. 
For instance, \citet{mehdizadeh2025epistemic} showed that Grokipedia citation behavior departs markedly from Wikipedia norms: it relies less on peer-reviewed academic sources and more heavily on user-generated or non-scholarly material, pointing to a reconfiguration of epistemic authority in AI-mediated knowledge production. 
Further, \citet{yasseri2025similar} conducts a comparative analysis from both textual and structural viewpoints. While Grokipedia articles often exhibit strong semantic alignment with their Wikipedia counterparts, they are generally longer, less lexically diverse, include fewer references per word, and display greater variability in structural organization.  Another analysis by \citet{triedman2025did} shows that much of Grokipedia content is heavily derived from Wikipedia, but its citation patterns include many more sources deemed unreliable by the Wikipedia community, especially on politically sensitive topics.
Complementing these content-level comparisons, \citet{coppolillo2025unexpected} examine the platforms search engine systems and find that both Wikipedia and Grokipedia often suggest results only loosely related to the user original query.
More broadly, prior work has examined bias in Wikipedia along dimensions such as gender representation \cite{wagner2015mans} and content quality dynamics \cite{ferrara2017dynamics}, while large-scale studies have developed methods for classifying political bias in online content \cite{weld2022political}.

\section{Methodology}

\paragraph{Controversial Topics.} Following \cite{coppolillo2025unmaskingconversationalbiasai}, we select six topics identified as among the most politically divisive issues in the United States, based on public opinion polls conducted in 2022, 2023, and 2024 by Gallup \cite{gallup} and the Pew Research Center \cite{pew, pew2}. 
Specifically, we analyze articles addressing: abortion, cannabis legalization, climate change, gender identity, gun control, and immigration. While many other controversial topics could be examined, we focus on these issues as representative of major contemporary American political cleavages.
% \todo[inline]{@Philipp Where/How did we take the data? Put links + some stats about them}

\paragraph{Data.}
We build a paired corpus of Grokipedia and Wikipedia articles covering the controversial topics described above.
% For each topic, we select the canonical English-language Wikipedia article and its Grokipedia counterpart.
Since Grokipedia adopts the same URL suffixes and titles as Wikipedia, we match articles by title/URL equivalence.
Both Grokipedia and Wikipedia pages were retrieved between November and December 2025. We crawl content via automated HTTP \texttt{GET} requests to the publicly accessible Grokipedia and Wikipedia websites. %, and the returned HTML was parsed to extract the main article body.
% No authentication or access restrictions were encountered during data collection.
To ensure cross-platform comparability, we remove non-prose and navigational content (including tables, reference lists, external links, and ``See also'' sections) from both platforms.
After preprocessing, the final dataset consists of 12 articles in total (6 per platform). 
We notice that, consistent with prior work~\cite{yasseri2025similar}, Grokipedia articles are substantially longer than their Wikipedia counterparts.
The mean (± standard deviation) article length is $11{,}105.8 \pm 2{,}859.3$ words for Grokipedia and $6{,}679.2 \pm 3{,}554.4$ words for Wikipedia.
%The corresponding median lengths are 11{,}626 words for Grokipedia and 6{,}947 words for Wikipedia.
%\todo[inline]{Check numbers}

\section{Results}
% \todo[inline]{@Philipp Add here the link for the source code in a github anonymous repository}
% \begin{figure}
%     \centering
%     \includegraphics[width=0.5\linewidth]{figures/legend_number_of_sections.pdf}
%     \\
%     \includegraphics[width=\linewidth]{figures/number_of_sections.pdf}
%     \caption{Number of sections on Grokipedia and Wikipedia, for each of the considered controversial topic.}
%     \label{fig:length}
% \end{figure}
\paragraph{Semantic Similarity.} 
Despite the aforementioned substantial difference in article length, the two platforms exhibit comparable structural granularity. Grokipedia articles contain an average of $27.8 \pm 6.0$ sections, compared to $26.5 \pm 13.6$ sections for Wikipedia articles. On this basis, we adopt a section-based semantic comparison strategy.

Specifically, for each pair of corresponding articles, we compute embeddings for all sections using GPT-5~\cite{openai_gpt5}. We then align sections across platforms by performing a best-match procedure: for each Grokipedia section, we identify the Wikipedia section with the highest embedding-based cosine similarity~\cite{church2017word2vec, gunawan2018implementation}.

\begin{figure}[h]

    \centering
    \includegraphics[width=0.8\linewidth]{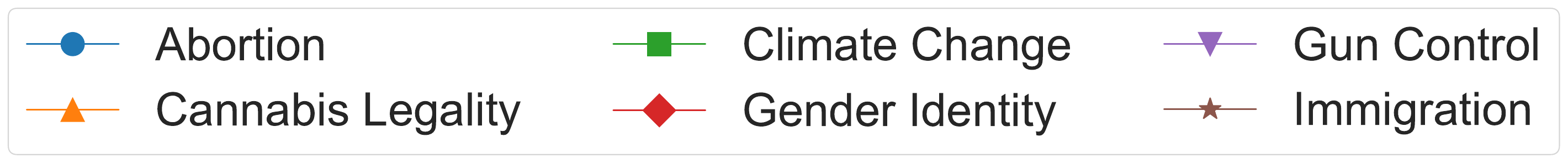}
    
    \centering
    \includegraphics[width=0.75\linewidth]{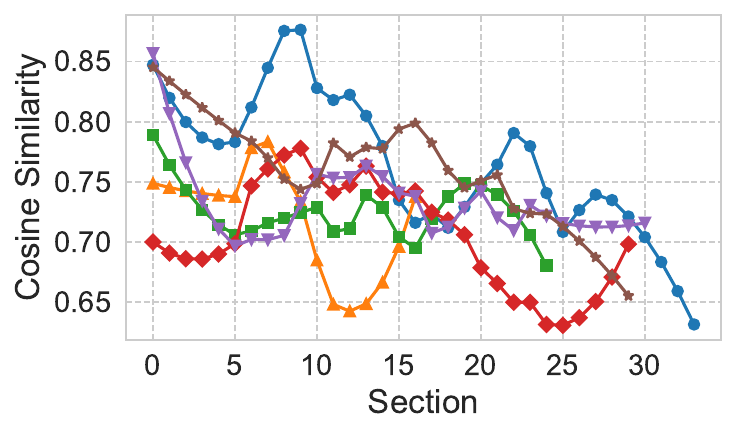}

    \caption{Section-level cosine similarity between Grokipedia and Wikipedia on controversial topics.}
    \label{fig:wikipedia_grokipedia_cosine_similarity}
\end{figure}

Figure \ref{fig:wikipedia_grokipedia_cosine_similarity} reports the section-level cosine similarity between Grokipedia sections and their best semantic match on Wikipedia, across the selected controversial topics. To improve readability, we apply the Savitzky-Golay filter~\cite{savitzky-filter} to smooth the curves. 
First, we observe a consistent decline in similarity as article sections progress: early sections exhibit relatively high semantic overlap between platforms, while  later sections show progressively lower similarity. This suggests increasing divergence in topical emphasis, narrative structure, or framing. 
While this pattern does not necessarily imply direct disagreement, it indicates that Grokipedia and Wikipedia increasingly depart in framing controversial issues beyond their core descriptions, even after accounting for differences in article length.

\begin{figure}[h]

    \centering
    \includegraphics[width=0.55\linewidth]{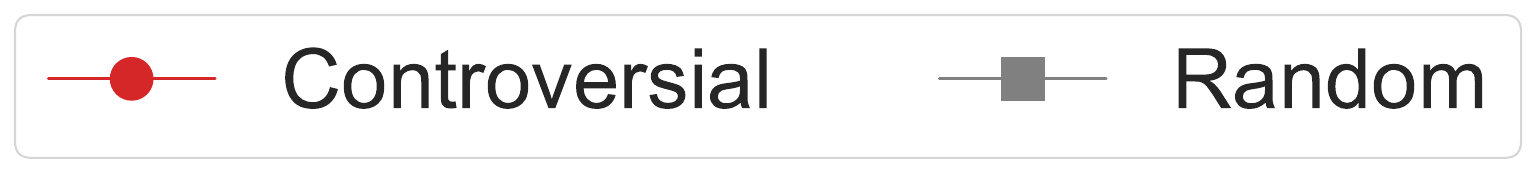}
    \centering
    \includegraphics[width=0.75\linewidth]{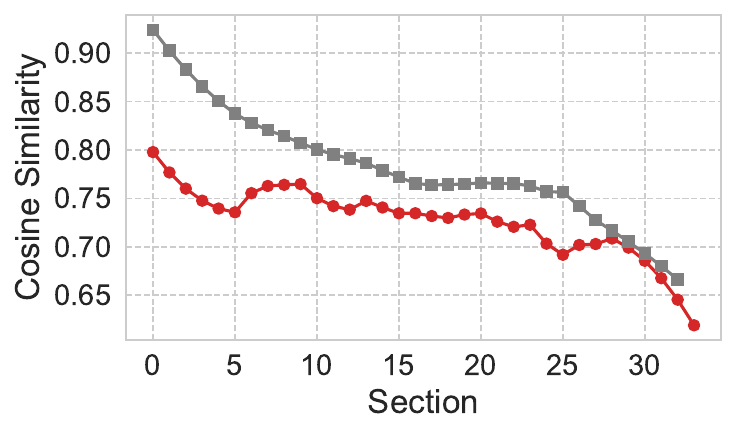}

    \caption{Average section-level cosine similarity between Grokipedia and Wikipedia for controversial pages compared to a random sample of 100 pages.}
    \label{fig:wikipedia_grokipedia_cosine_similarity_average}
\end{figure}

Using the same protocol, we further assess whether these discrepancies are specific to politically contested content.
To this end, we extend the cross-platform semantic similarity calculation to a random sample of $100$ articles that are present on both Grokipedia and Wikipedia but are not explicitly associated with controversial political topics.\footnote{https://en.wikipedia.org/wiki/Special:Random}.
The results of such comparison are shown in Figure~\ref{fig:wikipedia_grokipedia_cosine_similarity_average}. 
Interestingly, the randomly sampled article pairs exhibit substantially higher semantic similarity across the two platforms than the articles addressing controversial topics.

This finding suggests that political contestation plays a key role in shaping how Grokipedia and Wikipedia construct and frame content, leading to greater divergence on ideologically sensitive issues. 
% At the same time, we highlight that low semantic similarity is not always indicative of substantive disagreement. 
In some cases, differences arise from divergent article formats: some Wikipedia pages primarily consist of tables or structured lists, while Grokipedia presents largely narrative text.

These observations motivate our subsequent analysis, which moves beyond semantic overlap to explicitly examine the political orientation embedded in articles on the two platforms, allowing us to assess whether and to what extent ideological framing differs.

%\todo[inline]{@Philipp Add the baseline model description, GPT embeddings}

% \begin{figure}[t]

%     \centering
    
%     \includegraphics[width=\linewidth]{figures/cosine_similarity/baseline_pages.png}

%     \caption{}
%     \label{fig:baseline_pages}
% \end{figure}

\begin{figure}[h]

    \centering
    \begin{subfigure}[t]{0.26\textwidth}
        \centering
        \includegraphics[width=\linewidth]{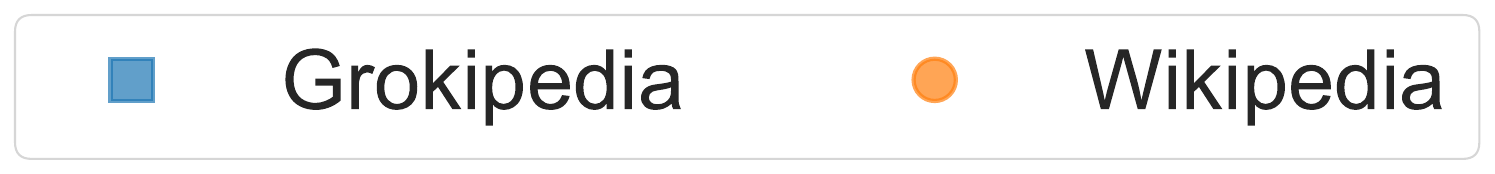}
    \end{subfigure}
    \centering
    \begin{subfigure}[t]{0.49\textwidth}
        \centering
        \includegraphics[width=\linewidth]{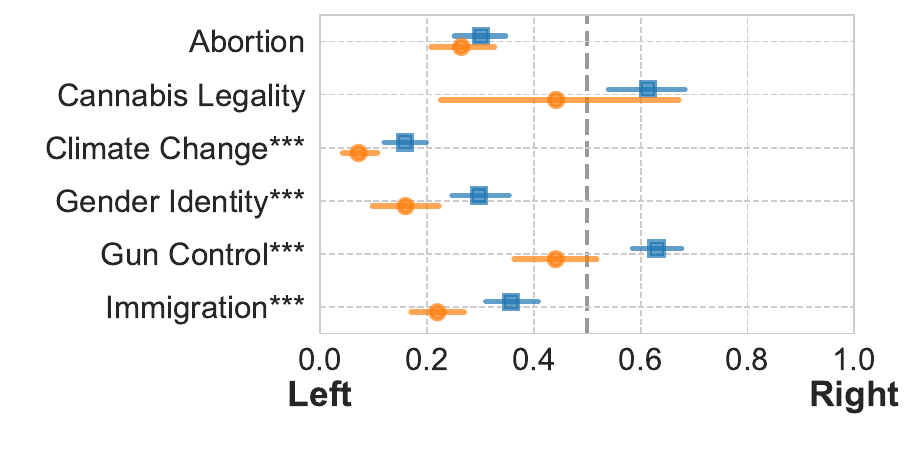}
    \end{subfigure}

    \caption{Political leaning on Grokipedia and Wikipedia across the considered controversial articles. Averages and $95\%$ confidence intervals are visualized.}
    \label{fig:wikipedia_grokipedia_political_distributions_mean_conf_int}
\end{figure}

\paragraph{Political Leaning.}
To classify the political stance of the selected controversial pages, we apply a RoBERTA-based classifier proposed in~\cite{espana2023multilingual}. The model is fine-tuned on a large corpus of newspaper articles ($\simeq 360,000$) to determine their political orientation, exhibiting strong performance ($\simeq 81\%$ accuracy on left-leaning articles, and $\simeq 89\%$ on right-leaning). 
For each text, it produces a score in $[0,1]$, where values close to $0$ (resp. $1$) indicate a stronger left- (resp. right-) wing orientation. 
Following the methodology in~\cite{espana2023multilingual}, we segment each page into individual sentences and apply the political stance classifier. 
We conduct a t-test~\cite{kim2015t} to assess whether the distributions differ significantly between Wikipedia and Grokipedia. 
The results are visualized in Fig.~\ref{fig:wikipedia_grokipedia_political_distributions_mean_conf_int}.
We find that, on average, most pages exhibit a left-leaning orientation on both Grokipedia and Wikipedia, with Grokipedia consistently scoring higher (i.e., being more right-leaning) across all topics. 
Additionally, on Cannabis Legality and Gun Control, Grokipedia displays an average political leaning closer to the right-leaning pole ($\simeq 0.6$). 

\begin{figure}[h]

    \centering
    \begin{subfigure}[t]{0.6\columnwidth}
        \centering
        \includegraphics[width=\linewidth]{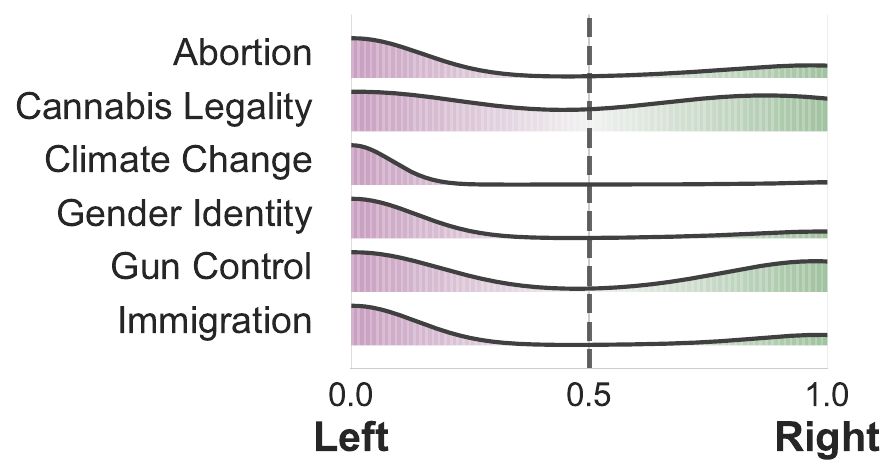}
        \caption{Wikipedia}
    \end{subfigure}
    \begin{subfigure}[t]{0.39\columnwidth}
        \centering
        \includegraphics[width=\linewidth]{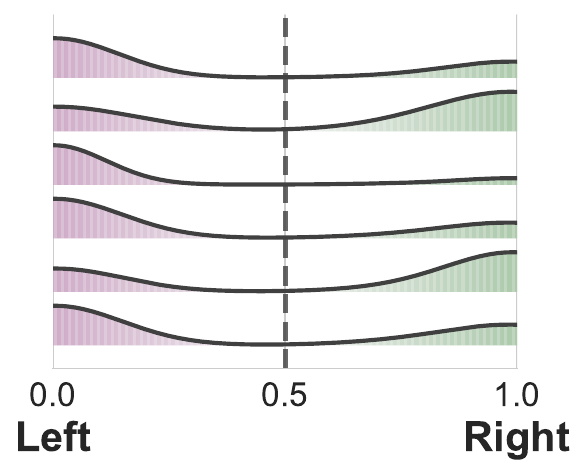}
        \caption{Grokipedia}
    \end{subfigure}

    \caption{Distribution of political leaning of controversial pages in (a) Wikipedia and (b) Grokipedia.}
    \label{fig:wikipedia_grokipedia_political_distributions}
    
\end{figure}

We further examine how the underlying political-leaning distribute for each controversial topic across platforms (Figure~\ref{fig:wikipedia_grokipedia_political_distributions}). 
We compute Hartigan’s Dip Test~\cite{hartigan1985dip} to assess bimodality. The results indicate that Grokipedia exhibits higher bimodality for Abortion, Gun Control, Immigration, and Climate Change ($p$-values $< 0.01$). 
Additionally, despite similar aggregate tendencies, Grokipedia exhibits a distribution with greater probability mass toward the right-leaning pole compared to Wikipedia.

\begin{figure*}[h]
    \centering
    \begin{subfigure}[t]{0.3\textwidth}
        \centering
        \includegraphics[width=\linewidth]{figures/political_stance/legend_political_leaning.pdf}
    \end{subfigure}
    \\
    \begin{subfigure}[b]{0.48\linewidth}
        \centering
        \includegraphics[width=\textwidth]{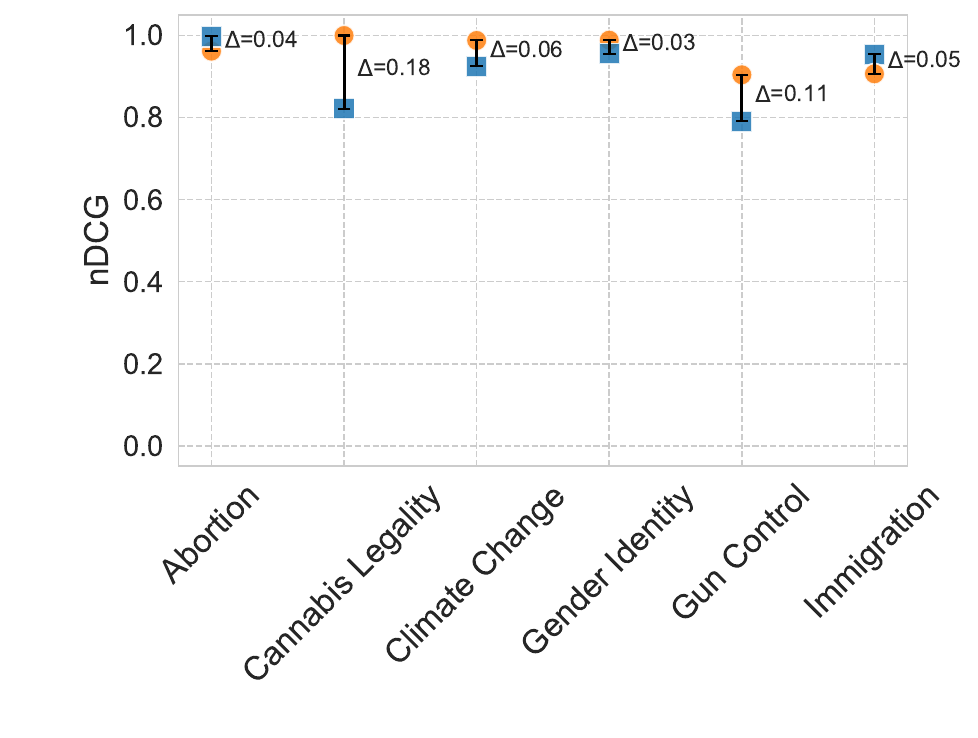}
        \caption{Left-leaning content}
    \end{subfigure}
    \begin{subfigure}[b]{0.48\linewidth}
        \centering
        \includegraphics[width=\textwidth]{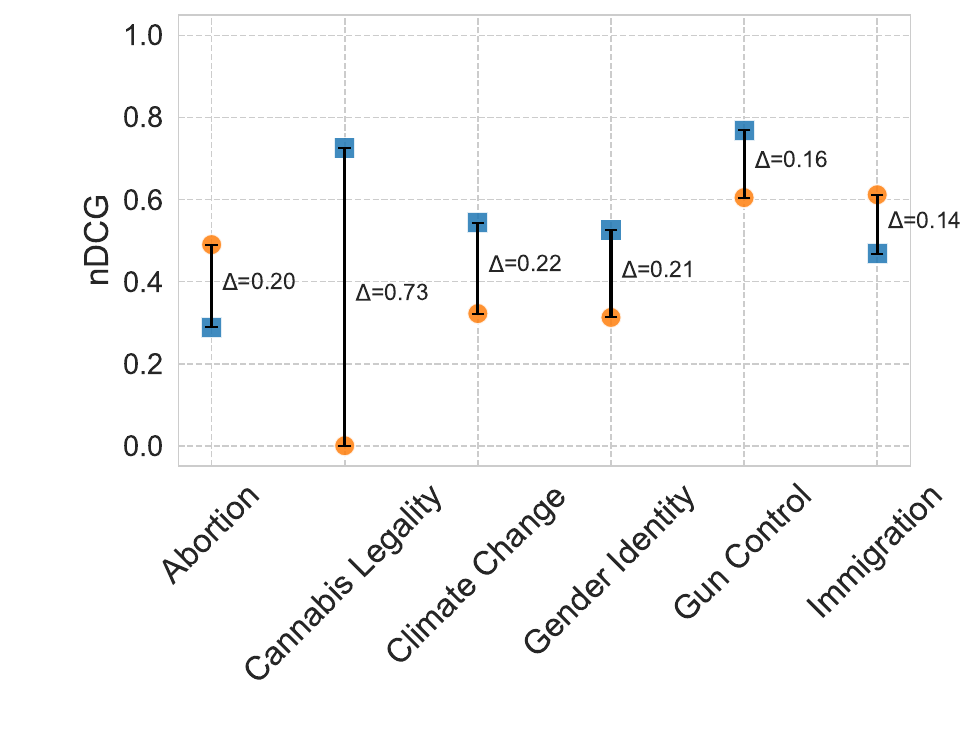}
        \caption{Right-leaning content}
    \end{subfigure}

    \caption{nDCG computed for (a) left- and (b) right-leaning content on Grokipedia and Wikipedia among controversial topics.}
    \label{fig:wikipedia_grokipedia_dcg}
\end{figure*}

% \begin{table}[ht]
% \centering
% \caption{Mean and standard deviation of the political leaning detected on Wikipedia and Grokipedia articles. Statistical comparison is performed under t-test (*** indicates a $p$-value $<0.01$).}
% \label{tab:wikipedia_grokipedia_statistical_test}
% \resizebox{0.8\columnwidth}{!}{
% \begin{tabular}{lccc}
% \toprule
% \textbf{Topic} & Wikipedia & Grokipedia \\
% \midrule
% $\text{Abortion}$               & $0.264 \pm 0.417$ & $0.301 \pm 0.448$ \\
% $\text{Cannabis Legality}$      & $0.442 \pm 0.456$ & $0.614 \pm 0.474$ \\
% $\text{Climate Change}^{***}$   & $0.072 \pm 0.249$ & $0.159 \pm 0.353$ \\
% $\text{Gender Identity}^{***}$  & $0.161 \pm 0.357$ & $0.297 \pm 0.443$ \\
% $\text{Gun Control}^{***}$      & $0.440 \pm 0.479$ & $0.629 \pm 0.468$ \\
% $\text{Immigration}^{***}$      & $0.219 \pm 0.404$ & $0.357 \pm 0.464$ \\
% \bottomrule
% \end{tabular}
% }
% \end{table}

Next, we investigate whether the two platforms systematically prioritize left- or right-leaning content in the top positions of their articles. This analysis is motivated by two key considerations: (i) user interactions with online content are strongly influenced by its position on the page~\cite{joachims2002optimizing, craswell2008experimental}; and (ii) users typically engage with online texts only partially, reading on average approximately $18\%$ of the words on a page~\cite{weinreich2008not}.

In light of these observations, we extract the first $18\%$ of words from each article and analyze the political leaning. 
We then employ the nDCG metric~\cite{wang2013theoretical, jarvelin2002cumulated} to quantify the extent to which left- and right-leaning sentences are preferentially ranked toward the top of the page. 
The metric evaluates the observed ranking against an ideal ordering with respect to a specific content type. 
%In our setting, we are interested in evaluating whether the platform places left-/right-leaning sentences at the very top of the article. 
Intuitively, higher scores indicate a stronger tendency to place that content type at the top of the page.

For each topic, we compute the nDCG gap (in absolute value) between the two platforms, denoted as $\Delta$. 
Figure~\ref{fig:wikipedia_grokipedia_dcg} depicts the results on prioritizing left- (a), and right-leaning content (b). 
First, we observe that Wikipedia tends to prioritize left-leaning content toward the top of the page more than Grokipedia for most of the topics (excluded Abortion and Immigration), while right-leaning content far less frequently appear in the top of the page. 
%Conversely, Grokipedia appears to give greater top-of-page prominence to left-leaning content for Abortion and Immigration (U.S.).
Second, the discrepancy between the platforms notably differ comparing left- and right-leaning content. 
Indeed, the $\Delta$ score is generally low across all topics when computed on left-leaning content ($\mu_\Delta \simeq 0.07, \sigma_\Delta \simeq 0.05$), while it considerably increases on right-leaning content ($\mu_\Delta \simeq 0.28, \sigma_\Delta \simeq 0.20$). 
This suggests that the two platforms similarly tend to high-rank left-leaning sentences in their articles, while they tend to locate right-leaning content in a considerable different way.
\begin{table}[ht]
\centering
\small
\resizebox{0.95\columnwidth}{!}{
\begin{tabular}{l p{0.38\columnwidth} p{0.38\columnwidth}}
\toprule
 & \textbf{Wikipedia} & \textbf{Grokipedia} \\
\midrule
\multirow{7}{*}{\rotatebox{90}{\textcolor{blue}{\textbf{Left}}}}
& \textit{A 2017 study found that waiting-period laws delaying firearm purchases by a few days reduce gun homicides by roughly 17\%.}
&\textit{ The Sandy Hook Elementary School shooting on December 14, 2012, resulted in 20 children and six adults killed, reigniting calls for federal action amid public outrage.} \\
\midrule
\multirow{10}{*}{\rotatebox{90}{\textcolor{red}{\textbf{Right}}}}
& \textit{Philosophy professor Michael Huemer argues that gun control may be morally wrong, even if effective, because individuals have a prima facie right to gun ownership for self-defense and recreation.}
& \textit{The Czech Republic amended its Charter of Fundamental Rights and Freedoms in 2021 to affirm the right to defend life with a weapon under a permit system, reinforcing self-defense as a basis for licensed ownership.} \\
\bottomrule
\end{tabular}
}
\caption{Phrases classified as left-/right- leaning, extracted from Wikipedia and Grokipedia on Gun Control.}
\label{tab:left_right_representative_phrases}
\end{table}

Finally, Table~\ref{tab:left_right_representative_phrases} presents anecdotal examples of sentences classified as left- and right-leaning from both platforms, with Gun Control selected as a representative topic.

\section{Conclusions}
This study offers a comparative analysis of how Wikipedia and Grokipedia cover well-established polarizing topics from a political perspective. Our results on semantic comparison reveal (i) a decaying semantic similarity in content presentation across corresponding articles, and (ii) more pronounced semantic differences on politically contested topics compared to randomly sampled ones. Further, the political leaning analysis indicates that both encyclopedias predominantly adopt left-leaning framings, whereas Grokipedia displays a more bimodal underlying distribution (more right-leaning oriented compared to Wikipedia).
Additionally, we find that both platforms similarly prioritize left-leaning content toward the top of their articles. However, Grokipedia tends to locate right-leaning content consistently higher on the page compared to Wikipedia. Taken together, these findings challenge the widespread perception of Grokipedia as an \textit{extreme} right-leaning encyclopedia, instead suggesting broadly comparable tendencies between the two platforms in their treatment of politically controversial topics, while still indicating a modest but consistent right-leaning bias in Grokipedia relative to Wikipedia.

\paragraph{Limitations.}
Our work presents several limitations that should be considered when interpreting the findings. First, our semantic similarity analysis relies on section-level embeddings and cosine similarity, which capture high-level semantic overlap but may overlook subtler rhetorical, tone or emphasis. Second, the analysis heavily relies on the quality of the political stance classifier adopted. While the selected model has been validated in prior work, political stance detection remains a challenging and inherently noisy task. Classifier predictions may be sensitive to lexical cues, context truncation, or domain-specific language, and may not fully capture nuanced or mixed ideological positions.

\paragraph{Future Work.}
This paper is amenable to future directions. Future work could extend this analysis to a broader and more diverse set of politically controversial topics, extending beyond U.S.-centric issues, to improve generalizability. Additionally, assessing factual accuracy, citation practices, and epistemic reliability would help contextualize observed differences in framing and political orientation. Finally, downstream user effects warrant investigation, particularly how differences in semantic framing and political leaning influence information retrieval, user trust, opinion formation, and polarization when these encyclopedias are accessed through search engines or AI assistants.

\section*{Acknowledgments}
This work has been partially funded by: (i) MUR on D.M.\ 351/2022, PNRR Ricerca, CUP H23C22000440007, (ii) MUR on D.M. 352/2022, PNRR Ricerca, CUP H23C22000550005.

%\clearpage

\bibliography{ref}

\end{document}